\documentclass[letterpaper, 10 pt, conference]{ieeeconf}  

\IEEEoverridecommandlockouts                              

\usepackage{times} 
\usepackage{amsmath} 
\usepackage{amssymb}  
\usepackage{graphicx}
\usepackage{dsfont}
\usepackage{color}
\usepackage{multicol}
\usepackage{url}
\usepackage{hyperref}
\usepackage{graphicx}

\usepackage{amsthm}

\newcommand{\bmtx}{\begin{bmatrix}}
\newcommand{\emtx}{\end{bmatrix}}
\newcommand{\bsmtx}{\left[ \begin{smallmatrix}} 
\newcommand{\esmtx}{\end{smallmatrix} \right]}
\newcommand{\R}{\mathbb{R}}

\usepackage{bm}  
\usepackage{graphicx} 
\usepackage{amsmath,amsthm,amssymb}

\usepackage{amsfonts}
\usepackage{mathtools}
\usepackage{amsmath}
\usepackage{amsmath, amssymb, amsthm} 
\usepackage{booktabs} 
\usepackage{mathrsfs}
\usepackage{booktabs}

\usepackage{algorithm}
\usepackage[noend]{algpseudocode}

\usepackage{siunitx}

\author{Turki Bin Mohaya \quad\quad Peter Seiler
	\thanks{T. Bin Mohaya and P. Seiler are with the Department of Electrical Engineering and Computer Science at the University of Michigan, Ann Arbor, MI 48109, USA. Email: \texttt{\small \{turki,pseiler\}@umich.edu}. The authors acknowledge funding from the Ford Motor Company.
    }
}

 \usepackage[style=ieee]{biblatex}
  
\AtBeginBibliography{\scriptsize}  
\date{April 2025}

\DeclareMathOperator*{\argmax}{arg\,max}

\usepackage{times}
\usepackage{psfrag,amsmath,amsfonts,verbatim,float}

\usepackage{url}
\usepackage{multicol}
\usepackage{float}

\usepackage[short]{optidef}
\usepackage{algorithm}
\usepackage[noend]{algpseudocode}

\usepackage{booktabs}
\usepackage{bbding}
\usepackage{pifont}

\usepackage{pgfplots}
\usepackage{tikz}
\usepackage{tikzscale}
\usepgfplotslibrary{groupplots}
\usetikzlibrary{backgrounds}
\pgfplotsset{compat=1.15}

\usepackage{xcolor}

\usepackage{soul}
\usepackage{caption}
\usepackage{tablefootnote}

\begin{document}


\title{\LARGE \bf Partial Attention in Deep Reinforcement \\ Learning for Safe Multi-Agent Control
}

\maketitle

\begin{abstract}

Attention mechanisms excel at learning sequential patterns by discriminating data based on relevance and importance. This provides state-of-the-art performance in today’s advanced generative artificial intelligence models. This paper applies this concept of an attention mechanism for multi-agent safe control. We specifically consider the design of a neural network to control autonomous vehicles in a highway merging scenario. The environment is modeled as a Decentralized Partially Observable Markov Decision Process (Dec-POMDP). Within a QMIX framework, we include partial attention for each autonomous vehicle, thus allowing each ego vehicle to focus on the most relevant neighboring vehicles. Moreover, we propose a comprehensive reward signal that considers the environment’s global objectives (e.g., safety and vehicle flow) and the individual interests of each agent. Simulations are conducted in the Simulation of Urban Mobility (SUMO). The results show better performance compared to other driving algorithms in terms of safety, driving speed, and reward.

\end{abstract}

\section{Introduction}
Highway merging is a fundamental yet challenging problem in autonomous driving. It requires agents to reason about dynamic interactions under uncertainty, where safe and efficient decisions depend not only on the agent's own state but also on the behaviors of surrounding vehicles. Traditional rule-based methods, although simple, often lack the flexibility to adapt to complex and dynamic merging situations. On the other hand, fully centralized deep reinforcement learning approaches struggle with scalability and are difficult to deploy in practical settings. Thus, decentralized policies that can selectively attend to relevant information present a promising direction for improving highway merging.

There are several similar works in the literature that deploy Multi-Agent Reinforcement Learning (MARL) \cite{yang2023towards, 10159552, zhou2022multi, du2025research, chen2025multi}. The authors in \cite{chen2025deep} proposed a simple recurrent unit to capture temporal patterns in the highway merging problem. These patterns are then fed to a Deep Deterministic Policy Gradient (DDPG) \cite{lillicrap2015continuous} network. After that, they enhance the training by introducing a prioritized replay buffer that samples experiences in proportion to the error of performance during that specific experience. This allows frequent replay of challenging scenarios for fast learning. The work in \cite{li2025multi} uses cross-attention mechanisms to fuse pose data and semantic data from different instruments on the vehicle, mainly for navigation. Their work shows good performance, but requires costly computational resources due to the complexity of the overall proposed framework. 

We consider the specific multi-agent scenario where vehicles are required to merge onto a highway. A model for this highway merging task is described in  Section~\ref{sec:probform}. Our proposed solution for autonomous merging builds on two ingredients that are reviewed in Section~\ref{sec:backgrnd}: attention mechanisms and QMIX \cite{rashid2020monotonic} for decentralized MARL. Section~\ref{sec:appr} then details our partial attention design and reward shaping. We illustrate our proposed method using the Simulation of Urban Mobility (SUMO) \cite{SUMO2018} and compare against other approaches (Section~\ref{sec:results}). Finally, Section~\ref{sec:conclusion} concludes the paper and discusses potential future directions.

Our contributions are twofold. First, we enhance the QMIX architecture by introducing a partial attention mechanism that focuses on the most critical interactions for each agent. Our notion of partial attention constitutes two elements: spatial attention and temporal attention. In spatial attention, we impose, by design, that each agent only observes the vehicle in front and the vehicle on the opposite merging road. In temporal attention, the neural network learns by itself to automatically focus on past time steps of these vehicles. This improves decision quality without incurring significant computational overhead. Second, we design a comprehensive reward structure that balances individual objectives, such as velocity maintenance and comfort, with global objectives such as collision avoidance and traffic flow improvement. Our method demonstrates improved safety and efficiency, validated through sophisticated simulations.

\section{Problem Formulation}
\label{sec:probform}

\subsection{Description}
\begin{figure*}
\centering
\includegraphics[height=4cm]{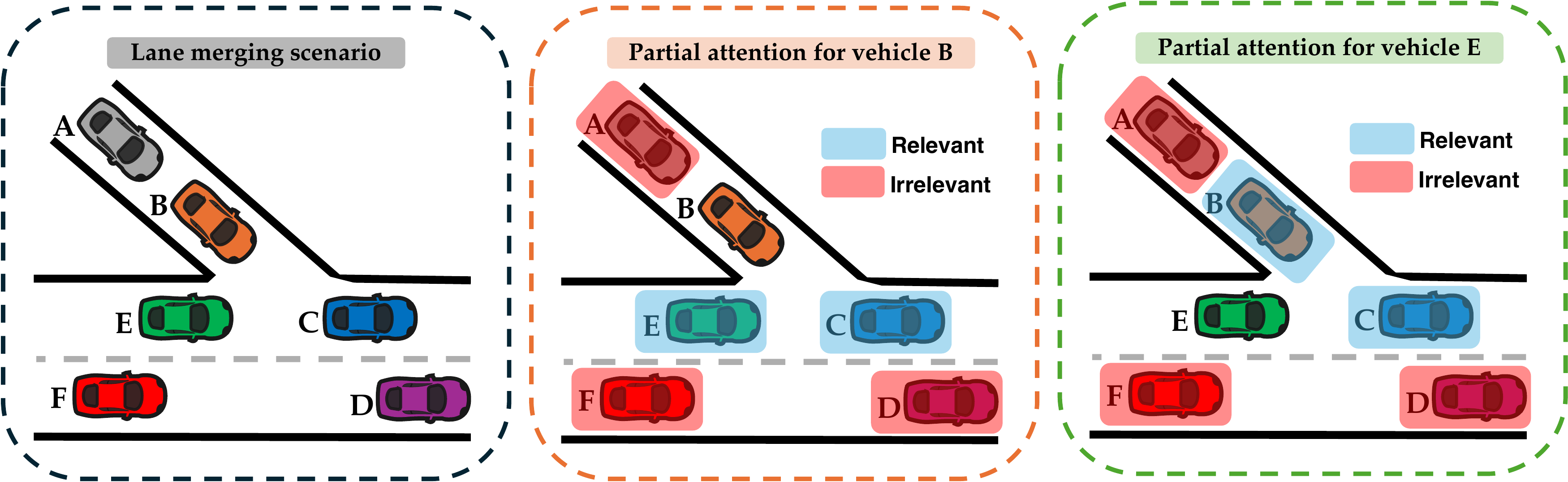}
\caption{{\textbf{Left:} The highway merging problem. \textbf{Middle} and \textbf{Right:} Our contribution is deploying partial attention to the most critical interactions for safe highway merging.}}
\label{fig:highway}
\vspace{-20pt}
\end{figure*}
The highway merging problem is challenging as agents with lower velocities approach other agents that drive with fast velocity profiles. Furthermore, it is safety-critical and typically requires a decentralized solution. \autoref{fig:highway} (left) illustrates the problem: vehicle B is merging while other highway vehicles are driving with higher velocities. The individual objectives of each agent mainly include safety (i.e., avoiding collision), and maintaining a desired velocity without altering the riders' comfort or compromising fuel efficiency. However, there are global objectives for the traffic that are sometimes in conflict with the objectives of individual vehicles. For example, it is desired to maintain a high average velocity to enhance vehicle flow, but highway agents may need to decelerate to allow merging. Also, the merging road should increase its throughput without negatively affecting the highway's average velocity.

Each agent makes decisions, in principle, based on all other nearby vehicles. However, not all other vehicles are equally relevant to an agent's decision. For example, in \autoref{fig:highway} (middle), vehicles A, F, and D are less relevant to the decision of vehicle B. On the other hand, vehicles C and E are of great importance to the merger decision. In fact, considering only information from these two agents can significantly reduce the computational cost for the merging decision-making process, and thus simplify the multi-agent highway merging problem. The dual is also true. In \autoref{fig:highway} (right), Vehicle E should focus on vehicles B and C while reducing attention to other vehicles.
\subsection{POMDPs}
The multi-vehicle environment is modeled as a Decentralized Partially Observable Markov Decision Process (Dec-POMDP)~\cite{amato2013decentralized} and represented by the following tuple:
\begin{align}
\mathcal{G}
=
\bigl(
  \mathcal{I},
  \mathcal{S},
  \{\mathcal{A}_i\},
  \mathcal{T},
  r,
  \{\Omega_i\},
  \mathcal{O},
  \gamma
\bigr).
\end{align}
The index set $\mathcal{I}:=\{1,\ldots,N\}$ labels the $N$ vehicles.  
A state $s\in\mathcal{S}$ summarizes the status of the highway. Each vehicle $i\in\mathcal{I}$ selects a discrete control $a_i\in\mathcal{A}_i$.  
Collecting the local controls yields the joint control
$a=(a_1,\dots,a_N)\in\mathcal{A}$, with
$\mathcal{A}=\times_{i\in\mathcal{I}}\mathcal{A}_i$, where $\times$ denotes the Cartesian product. State evolution is governed by the probability kernel
$\mathcal{T}:\mathcal{S}\times\mathcal{A}\times\mathcal{S}\rightarrow[0,1]$,
so that $\mathcal{T}(s,a,s')=\Pr(s'\mid s,a)$,
where $s'$ denotes the next state obtained after applying the action $a$ at state $s$. After the transition, vehicle $i$ receives a private observation
$o_i\in\Omega_i$.  The stacked observation
$o=(o_1,\dots,o_N)$ resides in $\Omega=\times_{i\in\mathcal{I}}\Omega_i$.  
A measurement function $\mathcal{O}:\mathcal{S}\times\mathcal{A}\times\Omega\rightarrow[0,1]$ defines the likelihood of an observation $o$ for a given state-action pair $(s,a)$ as: $\mathcal{O}(s,a,o)=\Pr(o\mid s,a)$.
Lastly, a scalar reward signal $r:\mathcal{S}\times\mathcal{A}\times\mathcal{S}\rightarrow\mathbb{R}$ provides the reward $r(s,a,s')$ accrued on a transition from $(s,a)$ to $s'$. Future scalar rewards are weighted by the discount factor $\gamma \in [0,1]$. Specifically, let $G_t$ denote the total discounted reward starting at time $t$ and going through the finite horizon $T$ of an episode. This total discounted reward is given by:
\begin{align}
G_t &= \sum_{k=t}^{T-1} \gamma^{k-t} r(k+1).
\label{eq:eq4}
\end{align}
Here, $r(k+1)$ is a shortened notation for the reward accrued at time $k+1$ for the transition from $(s(k),a(k))$ to the next state $s(k+1)$. The finite-horizon return can be equivalently expressed in recursive form as:
$G_t = r(t+1) + \gamma G_{t+1}, \quad G_T = 0,$
where $r(t+1)$ is the immediate reward and $\gamma G_{t+1}$ is the discounted future reward.  The terminal condition $G_T=0$ enforces the final horizon.

\section{Background}
\label{sec:backgrnd}
Deep neural networks~\cite{Goodfellow-et-al-2016} have shown an astonishing performance in learning complex patterns across vast dynamics and datasets. This is due to their ability to automatically extract features and optimize their weights through backpropagating the error loss. However, automatic feature extraction can, in some problems, suffer from an inability to focus on more relevant parts of the data. This can lead to sub-optimal solutions and inefficient training. Therefore, attention mechanisms were designed to equip neural networks with the ability to learn selective focus. This enables state-of-the-art performance in language translation and large language models~\cite{vaswani2017attention, chang2024survey}. We propose to apply similar attention mechanisms to autonomous driving.



\subsection{Attention Neural Networks}
Consider a vector-valued sequence $\{s_0, \, s_1, \ldots, s_T \} \subset \mathbb{R}^n$, and stack it row-wise to form the matrix $S := [s_0^\top; \, s_1^\top; \, \ldots; \, s_T^\top] \in \mathbb{R}^{(T+1) \times n}$. The scaled attention neural network~\cite{vaswani2017attention} takes $S$ as input and produces an output matrix $Z \in \R^{(T+1)\times d_{\text{o}}}$ where the dimension $d_{\text{o}}$ is described below.  The output is constructed from $H$ attention heads, each computing a query, key, and value. For head $j = 1, \ldots, H$, the query $\mathcal{Q}_j \in \mathbb{R}^{(T+1) \times d_j}$, key $\mathcal{K}_j \in \mathbb{R}^{(T+1) \times d_j}$, and value $\mathcal{V}_j \in \mathbb{R}^{(T+1) \times d_j}$ are computed as
\begin{align}
    \mathcal{Q}_j &= S W_Q^j, \quad
    \mathcal{K}_j = S W_K^j, \quad
    \mathcal{V}_j = S W_V^j,
\end{align}
where $W_Q^j, W_K^j, W_V^j \in \mathbb{R}^{n \times d_j}$ are the learned projection weights for the $j$-th head. Setting $H=1$ yields the single-head attention neural network, while setting $H>1$ employs multiple attention heads in parallel. Each head independently processes the input sequence, allowing the model to capture different aspects of the input data simultaneously. The attention weights for each head are
$\mathcal{A}_j = \texttt{Softmax}\!\left(\frac{\mathcal{Q}_j \mathcal{K}_j^\top}{\sqrt{d_j}}\right) \in \mathbb{R}^{(T+1) \times (T+1)}$,
where $d_j$ is the dimension of the keys and queries. The output of head $j$ is then
$Z_j = \mathcal{A}_j \mathcal{V}_j \in \mathbb{R}^{(T+1) \times d_j}$.
The outputs of all heads are concatenated and linearly transformed to produce the final output matrix of the attention neural network:
\begin{align}
Z &= 
\begin{bmatrix}
Z_1 & Z_2 & \dots & Z_H
\end{bmatrix}
W_O \in \mathbb{R}^{(T+1) \times d_{\text{o}}},
\end{align}
where $d=\sum_{j=1}^H d_j$ is the column dimension of the concatenated matrix.  Moreover, $W_O \in \R^{d \times d_o}$ is a learned projection matrix that maps the combined output to the desired output dimension $d_{\text{o}}$. This multi-head approach enables the model to capture diverse patterns and relationships within the input sequence. This enhances its ability to learn complex dependencies and improve overall performance.

\subsection{QMIX}

Multi-agent deep reinforcement learning (MADRL) provides a robust method to learn practical policies or controllers for stochastic multi-agent environments. This is due to their ability to capture highly non-linear stochastic dynamics by iteratively interacting with the environment and observing the consequences. Independent Q-learning (IQL)~\cite{tan1993multi} decomposes the multi-agent problem into parallel single-agent problems that are simultaneously collocated. In IQL, each agent assumes the other agents are non-moving in the environment. However, learning for the agents must be coordinated to address the nonstationary environment.  

The principle of optimality \cite{sutton18} can be used to express a recursive (centralized) solution for the total action-value function $Q_{tot}$ of all agents.
QMIX~\cite{rashid2020monotonic} approximates the total action-value function through a nonlinear mapping $\texttt{Mix}(\cdot)$ of their individual action-value functions $\{Q_i\}_{i=1}^N$ of the $N$ agents. This nonlinear mapping is implemented as a neural network and is given by
\begin{align}
\label{eq:QMIXQtot}
Q_{tot}(o, a; \theta) = \texttt{Mix}(o, a, &Q_1(o_1, a_1; \theta_1), \dots \nonumber  \\
&Q_N(o_N, a_N; \theta_N), \theta_{N+1}),
\end{align}
where $\theta$ includes the weights of the individual action-value functions $\{\theta_i\}_{i=1}^N$ and the weights of the mixing network $\theta_{N+1}$.

QMIX minimizes a Deep Q-Network~\cite{mnih2015human} regression loss over a replay mini‑batch of size $B$. To define this loss, let $o(m)$ and $a(m)$ denote the joint observation and the joint action at sample $m$, respectively. The one-step target is
\begin{align}
  y(m)
  =
  r(m)
  +
  \gamma
  \max_{ a'}
  Q_{\mathrm{tot}}\bigl(o'(m), a';\theta^{-}\bigr),
  \label{eq:qmix_target}
\end{align}
where $r(m)$ is the immediate reward, $o'(m)$ is the next joint observation at sample $m$, $a'$ is the next joint action, and $\theta^{-}$ are the parameters of a slowly updated target network.
The loss over a replay mini‑batch of size $B$ is then given by
\begin{align}
  \mathcal{L}(\theta)\;=\;\frac1B\sum_{m=1}^{B}
  \bigl[y(m) - Q_{\mathrm{tot}}(o(m), a(m);\theta)\bigr]^{2}.
  \label{eq:qmix_loss}
\end{align}
  
The mixing network weights are constrained to be non‑negative.  
This architectural choice enforces the monotonicity condition
$  \frac{
    \partial
    Q_{\mathrm{tot}}(o,a)
  }{
    \partial
    Q_i(o_i,a_i)
  }
  \ge 0,
  \forall i\in\mathcal I,$
ensuring that an improvement in any agent’s value cannot reduce the global estimate. As a result of $Q_{\mathrm{tot}}$ being monotone, the joint maximizer factorizes into independent maximizers. This is due to the Individual-Global-Max (IGM) principle~\cite{son2019qtran}:
\begin{align}
  \argmax_{a}\,
  Q_{\mathrm{tot}}(o,a)
  =
  \begin{bmatrix}
    \argmax_{a_1} Q_1(o_1,a_1) \\
    \vdots \\
    \argmax_{a_N} Q_N(o_N,a_N)
  \end{bmatrix},
  \label{eq:igm}
\end{align}
so team‑optimal actions can be obtained by greedy choices made locally by each agent.

During learning, the full joint trajectory $o$ is available to the mixing network, providing a centralized viewpoint that removes the non-stationarity that is a consequence of purely decentralized training. In this phase, each agent selects a random action $a_i^{\text{random}}$ with probability $\epsilon$ to encourage exploration, while with probability $1-\epsilon$ it selects the greedy action $a_i^\star = \arg\max_{a_i} Q_i(o_i, a_i)$. Formally, the individual action selection at each step is
\begin{align}
a_i = 
\begin{cases}
a_i^{\text{random}} & \text{with probability } \epsilon, \\
a_i^\star & \text{with probability } 1-\epsilon.
\end{cases}
\end{align}
The exploration probability $\epsilon$ decays over time according to a fixed decay rate $\epsilon_{\text{decay}}$ until reaching a specified minimum value $\epsilon_{\text{min}}$. During evaluation, the model fully exploits its learned policy by setting $\epsilon=0$, thereby always choosing the greedy action.
After convergence, the mixing network is discarded, and each agent selects its action as
$a_i^{\star}$ using local information $o_i$.  
It is worth noting that even with each agent selecting actions based on local information, \eqref{eq:igm} guarantees that the resulting joint action remains optimal.


\section{Approach}
\label{sec:appr}

\subsection{Agent State Design}

The state of each autonomous vehicle (AV) is designed to encapsulate both its own dynamics and the dynamics of its immediate surroundings through the partial attention mechanism. This approach allows the $i$-th AV to focus on the most relevant interactions, specifically the vehicle directly ahead of it, $\text{f}(i)$, and the vehicle approaching from the opposite road, $\text{o}(i)$. To illustrate, vehicle E in \autoref{fig:highway} (right) has a front vehicle C and an opposite vehicle B. On the other hand, in \autoref{fig:highway} (middle), vehicle B has the front vehicle C and the opposite vehicle E. Similarly, vehicle A has the front vehicle B and the opposite vehicle E. Finally, vehicles D and C have neither front nor opposite vehicle. Every agent's state representation is composed of its current state and the historical states of these two neighboring vehicles over a specified time window. The \emph{kinematic state} of vehicle $ i $ at time $ t $ is defined as
$\bar{S}_i(t) = \begin{bmatrix} x_i(t) & y_i(t) & v_i(t) & a_i(t) \end{bmatrix}^\top \in \mathbb{R}^4$,
where $ x_i(t) $ and $ y_i(t) $ are the position coordinates, $ v_i(t) $ is the speed, and $ a_i(t) $ is the acceleration of the vehicle.

The \emph{information state} of vehicle $ i $ at time $ t $, denoted $ S_i(t) \in \mathbb{R}^{4 + 8(w+1)} $, is defined as
\begin{align}
S_i(t) = \begin{bmatrix} \bar{S}_i(t)^\top & \mathrm{vec}(F_i(t))^\top & \mathrm{vec}(O_i(t))^\top \end{bmatrix}^\top, \label{eq:agent_state}
\end{align}
where $\mathrm{vec}(\cdot)$ denotes row-major vectorization of a matrix into a flat vector, and $ F_i(t), O_i(t) \in \mathbb{R}^{(w+1) \times 4} $ represent the historical kinematic states of the front and opposite vehicles over the most recent $ w + 1 $ time steps:
\begin{align}
F_i(t) &= \begin{bmatrix} \bar{S}_{\text{f}(i)}(t-w) & \cdots & \bar{S}_{\text{f}(i)}(t) \end{bmatrix}^\top, \\
O_i(t) &= \begin{bmatrix} \bar{S}_{\text{o}(i)}(t-w) & \cdots & \bar{S}_{\text{o}(i)}(t) \end{bmatrix}^\top.
\end{align}
Here, $ \bar{S}_{\text{f}(i)}(t) $ and $ \bar{S}_{\text{o}(i)}(t) $ denote the kinematic states of the front and opposite vehicles associated with agent $ i $, respectively. We assume that each agent can perfectly observe the kinematic state of its front and opposite vehicles. This corresponds to an idealized vehicle-to-vehicle (V2V) communication or sensing model.

\subsection{Partial Attention}

The front and opposite vehicle histories, $F_i(t)$ and $O_i(t)$, are leveraged by processing through a multi-head attention mechanism. This processing captures temporal dependencies and contextual interactions. This constructs a comprehensive state representation for each agent. The design begins with layer normalization~\cite{ba2016layernormalization}. Given a vector $x$, the output $y$ of the layer normalization operation can be expressed as
\begin{align}
    y = \frac{x-E[x]}{\sqrt{Var[x]+\rho}}\odot\kappa+\beta,
\end{align}
where $E[x]$ is the mean of $x$, $Var[x]$ is the variance of $x$, and $\odot$ represents element-wise multiplication. We set $\rho = 10^{-5}$ while $\kappa$ and $\beta$ are affine learnable parameters. We utilize PyTorch~\cite{Ansel2024} to conduct this operation. The deployment of this function is discussed next.

Each historical sequence is first normalized to ensure consistent scaling across different features and time steps:
\begin{align}
\tilde{F}_i(t) = \texttt{LN}(F_i(t)), \qquad
\tilde{O}_i(t) = \texttt{LN}(O_i(t)),
\end{align}
where $ \texttt{LN}(\cdot) $ denotes layer normalization applied to each feature vector within the sequence. Following normalization, each token in the sequence is projected into a higher-dimensional embedding space to enhance the capacity of the attention mechanism:
\begin{align}
X_{F_i}(t) &= \tilde{F}_i(t)\, W_{F}^\top + \mathbf{1}_{w+1}b_{F}^\top \in \mathbb{R}^{(w+1) \times d_{\text{model}}}, \\
X_{O_i}(t) &= \tilde{O}_i(t)\, W_{O}^\top +\mathbf{1}_{w+1}b_{O}^\top \in \mathbb{R}^{(w+1) \times d_{\text{model}}},
\end{align}
where
$W_{F}, W_{O} \in \mathbb{R}^{d_{\text{model}} \times 4}$ and $b_{F}, b_{O} \in \mathbb{R}^{d_{\text{model}}}$
denote the weight matrices and bias vectors of the corresponding linear transformations, and $\mathbf{1}_{w+1}\in \mathbb{R}^{w+1}$ is a vector whose entries are all ones. Here, $ d_{\text{model}} $ is the dimensionality of the embedding space. Next, the projected sequences $ X_{F_i}(t) $ and $ X_{O_i}(t) $ serve as inputs to separate multi-head attention modules tailored for the front and opposite vehicle histories:
\begin{align}
Z_{F_i}(t) &= \texttt{MHA}(X_{F_i}(t)) \in \mathbb{R}^{(w+1) \times d_{\text{model}}},
\label{eq:26}\\
Z_{O_i}(t) &= \texttt{MHA}(X_{O_i}(t)) \in \mathbb{R}^{(w+1) \times d_{\text{model}}}.
\label{eq:27}
\end{align}
The function $ \texttt{MHA}(\cdot) $ denotes the multi-head attention operation on the input sequence. This mechanism allows the model to focus on different parts of the input sequences simultaneously, capturing diverse aspects of temporal dependencies.

To synthesize the information captured by the attention mechanism, a weighted aggregation of the attention outputs is performed along the temporal axis. This aggregation gives more emphasis on the most recent historical data, reflecting their higher relevance in the current decision-making context. We define $\alpha = \texttt{Softmax}( \Psi_{0.5,1} ) \in \mathbb{R}^{w+1}$ where $\texttt{Softmax}(\cdot)$ normalizes the vector $\Psi_{0.5,1}$, which is an increasing evenly spaced vector between 0.5 and 1.0 over the $w+1$ time steps. Then, the weighted embeddings are
\begin{align}
E_{F_i}(t) = Z_{F_i}(t)^\top\alpha,\qquad
E_{O_i}(t) = Z_{O_i}(t)^\top\alpha.
\end{align}
This contracts the temporal axis of $Z_{F_i}(t)$ and $Z_{O_i}(t)$, producing a single embedding vector that weights the relevance of each time step. To further enhance the expressive power of the embeddings $ E_{F_i}(t) $ and $ E_{O_i}(t) $, they are passed through two feed-forward neural networks (FFN) with residual connections and layer normalization as
\begin{align}
E'_{F_i}(t) &= \texttt{LN}\left( E_{F_i}(t) + \texttt{FFN}_{\texttt{F}}(E_{F_i}(t)) \right) \in \mathbb{R}^{d_{\text{model}}}, \\
E'_{O_i}(t) &= \texttt{LN}\left( E_{O_i}(t) + \texttt{FFN}_{\texttt{O}}(E_{O_i}(t)) \right) \in \mathbb{R}^{d_{\text{model}}},
\end{align}
where $ \texttt{FFN}_{\texttt{F}}(\cdot) $ and $ \texttt{FFN}_{\texttt{O}}(\cdot) $ are feed-forward networks comprising linear transformations and non-linear activations. The residual connections facilitate gradient flow and stabilize training by allowing the model to retain information from earlier layers.

The final state representation $ S'_i(t) $ for agent $ i $ at time step $ t $ is constructed by concatenating the agent's own state vector $ \bar{S}_i(t) $ with the aggregated embeddings from the front and opposite vehicle histories:
\begin{align}
S'_i(t) = \begin{bmatrix} \bar{S}_i(t)^\top &  E'_{F_i}(t)^\top & E'_{O_i}(t)^\top \end{bmatrix}^\top \in \mathbb{R}^{4 + 2d_{\text{model}}}.
\end{align}
This enriched state vector integrates both the intrinsic dynamics of the agent and the contextual information derived from its immediate vehicular environment through partial attention. The comprehensive state $ S'_i(t) $ is subsequently fed into the QMIX utility network, which leverages this information to estimate the utility values for each possible action, thereby facilitating coordinated and efficient decision-making during merging maneuvers.

\subsection{Reward Design}
The reward structure is designed to incentivize desirable behaviors and penalize undesirable actions. It integrates both global and local reward components, ensuring that individual agent performance aligns with overall traffic efficiency and safety objectives. The sum of all terms gives a comprehensive reward signal that facilitates effective learning of the QMIX model and enables coordinated decision-making in the merging scenario.

\textit{Global Reward Signal:} The global reward is designed to promote collective traffic efficiency and safety through metrics that reflect the overall performance of all agents. First, to promote safety, a penalty is imposed when a collision occurs via the term
\begin{align}
r_{\text{collision}}(t) = -c_1 N_{\text{collision}}(t),
\end{align}
where $ N_{\text{collision}}(t) $ represents the number of collisions at time step $ t $, and $ c_1 $ is a positive weighting coefficient. Upon receiving this penalty, the episode is terminated. Next, to maintain efficient traffic flow and minimize congestion, a traffic flow term encourages agents to sustain desired speeds:
\begin{align}
r_{\text{flow}}(t) = c_2 \overline{v}_{\text{highway}}(t) + c_3 \overline{v}_{\text{merging}}(t),
\end{align}
where $ \overline{v}_{\text{highway}}(t) $ and $ \overline{v}_{\text{merging}}(t) $ denote the average velocities on the highway and merging lanes at time step $ t $, and $ c_2, c_3 $ are positive balancing coefficients. To mitigate idling and reduce travel time, we define the waiting penalty as
\begin{align}
r_{\text{waiting}}(t) = -c_4 T_{\text{waiting}}(t),
\end{align}
where $ T_{\text{waiting}}(t) $ is the cumulative time vehicles spend below the minimum allowed velocity, and $ c_4 $ is the weighting coefficient. Finally, to encourage route completion, we define
\begin{align}
r_{\text{goal}}(t) = c_5 N_{\text{goal}}(t),
\end{align}
where $ N_{\text{goal}}(t) $ denotes the number of vehicles that successfully reached their destinations at time step $ t $, and $ c_5 $ is the corresponding weighting coefficient.

\textit{Individual Reward Signal:} In addition to the global reward, individual rewards are directly attributed to each agent based on its specific actions and states, ensuring independent performance maximization. Each agent is rewarded for tracking its desired velocity via
\begin{align}
r_{\text{velocity}, i}(t) = -c_6 \frac{\left| v_i(t) - v_{i,\text{desired}}\right|}{v_{i,\text{desired}}},
\end{align}
where $ v_i(t) $ is the velocity of agent $ i $ at time $t$, $ v_{i,\text{desired}} $ is its target velocity, and $ c_6 $ is a positive weighting coefficient. Fuel-efficient driving is promoted through
\begin{align}
r_{\text{efficiency}, i}(t) = -c_7 \text{Fuel}_i(t),
\end{align}
where $ \text{Fuel}_i(t) $ is the fuel consumption of agent $ i $ at time step $ t $, and $ c_7 $ is the weighting coefficient. Finally, passenger comfort is enforced by penalizing sharp accelerations via
\begin{align}
r_{\text{comfort}, i}(t) = -c_8 \left| a_i(t) \right|,
\end{align}
where $ a_i(t) $ is the acceleration of agent $ i $ at time step $ t $, and $ c_8 $ is the corresponding weighting coefficient.
\begin{figure*}
    \centering
    \includegraphics[width=0.9\linewidth]{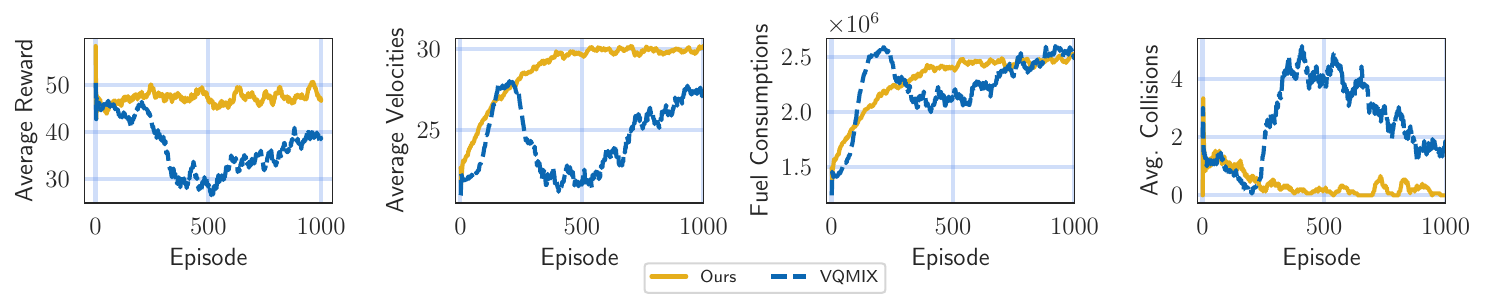}
    \caption{Performance during the training phase.}
    \label{fig:training}
    \vspace{-0.15in}
\end{figure*}
\begin{figure*}
    \centering
    \includegraphics[width=0.9\linewidth]{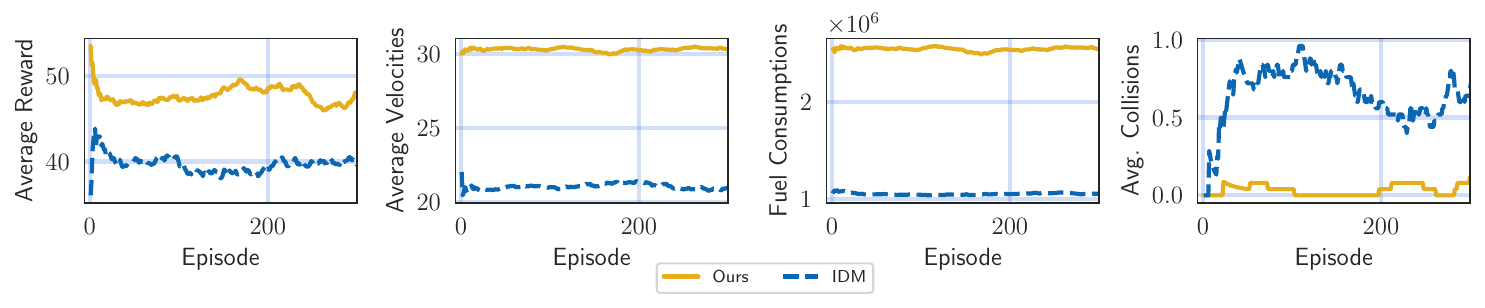}
    \caption{Comparison of the proposed method against SUMO IDM in the evaluation phase.}
    \label{fig:plt}
        \vspace{-.2in}
\end{figure*}
\textit{Final Reward Signal:} The comprehensive reward $r(t)$ at step $t$ is the sum of all global and individual agent reward terms:
\begin{align}
r(t) &= r_{\text{collision}}(t) + r_{\text{flow}}(t) + r_{\text{waiting}}(t) + r_{\text{goal}}(t) \nonumber \\
&\quad + \sum_{i=1}^N r_{\text{velocity}, i}(t) + r_{\text{efficiency}, i}(t) + r_{\text{comfort}, i}(t).
\end{align}
The total reward $r(t)$ enters the discounted return defined in \eqref{eq:eq4}. This reward structure ensures that agents are simultaneously motivated to maintain safe and efficient driving, both individually and collectively, enhancing overall traffic dynamics during highway merging.

\section{Results}
\label{sec:results}

We utilize the Simulation of Urban Mobility (SUMO)~\cite{SUMO2018} to build the highway environment. Neural networks are designed and trained via PyTorch~\cite{Ansel2024} and then deployed in SUMO through the integration of TraCI~\cite{wegener2008traci}. Our model is trained for $1000$ episodes, each with a maximum of $1000$ time steps. The hyperparameters used to train the networks are summarized in \autoref{tab:hyperparams}.

The highway has two lanes and a length of $400$ meters, but vehicles are not allowed to change lanes. The merging road has a length of $100$ meters and one lane. \autoref{tab:params-sumo} reports the random parameters used to generate vehicles for each episode. This includes a fixed number of agents, the initial velocities of highway and merging road vehicles, the assigned departure time (the time at which the vehicle begins to exist), and the assigned road, i.e., label $HW$ for highway spawn and $M$ for a merging road spawn. 
\begin{table}
    \scriptsize
    \centering
    \caption{Training Hyperparameters}
    \begin{tabular}{c|c}
        \hline
        \textbf{Parameter} & \textbf{Value} \\ \hline
        Episodes & $1000$ \\
        Maximum time steps per episode & $1000$ \\
        Optimizer & AdamW~\cite{loshchilov2018decoupled} \\
        $B$ (batch size) & $256$ \\
        $\gamma$ (discount factor) & $0.99$ \\
        Learning rate & $0.0001$ \\
        $\epsilon$ & $1.0$ \\
        $\epsilon_{\text{min}}$ & $0.05$ \\
        $\epsilon_{\text{decay}}$ & $0.99$ \\
        Target network update interval & $4$ episodes \\
        TraCI sampling interval ($\mathrm{s}$) & ${0.1}$ \\
        Replay buffer size & $1000000$ \\
        Action space - acceleration (${\mathrm{m}}/{\mathrm{s}^{2}}$) & $[-6, -3, -2, -1,0,1, 2, 3, 6]$ \\
        $w$ & $9$ \\
        \hline
    \end{tabular}
    \label{tab:hyperparams}
\end{table}
\begin{table}
    \scriptsize
    \centering
    \caption{Vehicle Generation Parameters}
    \begin{tabular}{c|c}
        \hline
        \textbf{Parameter} & \textbf{Value} \\ \hline
        Number of agents & $16$ \\
        Initial velocity for highway agents ($\mathrm{m}/\mathrm{s}$) & $\sim \text{Uniform}([7, 10])$ \\
        Initial velocity merging agents ($\mathrm{m}/\mathrm{s}$) & $\sim \text{Uniform}([4, 8])$ \\
        Departure time ($\mathrm{s}$) & $\sim \text{Uniform}([0, 100])$ \\
        Route & $\sim \text{Uniform}\left(\{HW, M\}\right)$ \\
        Vehicle length ($\mathrm{m}$) & $5.0$ \\
        \hline
    \end{tabular}
    \label{tab:params-sumo}
    \vspace{-0.15in}
\end{table}

\autoref{tab:params-reward} reports the reward coefficients. $c_1$ is large to critically prioritize vehicle safety. $c_2$ and $c_3$ are balanced to enhance both the merging flow and the highway flow. $c_4$ penalizes waiting time without overshadowing other terms. $c_5$ rewards safely reaching the goal, but it is not enough to receive a large reward. $c_6$ increases the vehicle's velocity, while $c_7$ is designed not to exceed high speeds. Lastly, the comfort term $c_8$ enables smoother acceleration signals.
\begin{table}
\vspace{-.1in}
    \scriptsize
    \centering
    \caption{Reward Coefficients}
    \begin{tabular}{c|c|c|c|c|c|c|c|c}
        \hline
        \textbf{Coefficient} & $c_1$ &  $c_2$ &  $c_3$ &  $c_4$ &  $c_5$ &  $c_6$ &  $c_7$ &  $c_8$  \\ \hline
        \textbf{Value} & 40 & 0.5 & 0.9 & 1.0 & 1.0 & 3.0 & 0.00001 & 0.01 \\
        \hline
    \end{tabular}
    \label{tab:params-reward}
\end{table}

The training was conducted on a MacBook Pro equipped with an Apple M4 chip, and it lasted approximately 56 minutes. \autoref{fig:training} shows the performance of the proposed method during the training phase. Within 500 episodes, the model converged to high average velocities while significantly reducing the average number of collisions. Fuel consumption increases during training as a result of increasing the average velocities of vehicles. Furthermore, we conduct an ablation study by training the model without the temporal attention layers (i.e, omitting \eqref{eq:26} and \eqref{eq:27}). This ablation comparison is referred to as Vanilla QMIX (VQMIX) in \autoref{fig:training}. Its performance diverges within 300 episodes of training. This illustrates the importance of the temporal attention layers to focus on the critical temporal dynamics. Hence, the ablation study demonstrates the effectiveness of the proposed method. 

Next, we discuss the results during the evaluation phase. \autoref{fig:plt} compares our Partial Attention QMIX model with SUMO's Intelligent Driving Model (IDM)~\cite{treiber2000congested} in four performance metrics. In terms of average reward, our method shows a clear improvement over episodes. For average velocities, our model reaches and maintains a higher level, meaning vehicles move more smoothly and efficiently. Fuel consumption is slightly higher with our approach, probably due to higher velocity. Finally, the number of collisions drops significantly during training with our model, but IDM continues to have more frequent crashes. These results illustrate how carefully tailoring the information state of the ego vehicle to include only important nearby interplay dynamics can enhance training speed and evaluation performance. 


\section{Conclusion}
\label{sec:conclusion}
This paper presented a highway merging framework integrating partial attention into QMIX, where attention operates on two complementary levels: spatial attention identifies the relevant vehicles, while temporal attention extracts informative past states. Combined with a hybrid reward signal balancing global and individual objectives, this design yields safer and more efficient merging behavior. Simulation results in a SUMO-based environment confirm significant improvements in collision rate, average velocity, and overall reward over a standard driving baseline, with the trade-off of higher fuel consumption driven by increased speeds. Future work will extend the framework to multi-lane and mixed-autonomy settings where autonomous and human-driven vehicles coexist.


\printbibliography
\end{document}